\documentclass[fleqn,twoside]{article}%
\usepackage{amsmath}
\usepackage{amsfonts}
\usepackage{amssymb}
\usepackage{graphicx}
\usepackage{cite}
\def\be{\begin{equation}}
\def\ee{\end{equation}}
\def\bi{\bibitem}
\begin{document}
\title{Viscoelastic flow past an infinite plate with suction and constant heat flux.}
\author{Abhik Kumar Sanyal$^1$ and D. Ray $^2$}
\maketitle
\noindent
\begin{center}
\noindent
$^1$ Dept of Physics, University College of Science,\\
92 A.P.C. Road, Calcutta-700009, India.\\
$^2$ Dept of Applied Mathematics, University College of Science,\\
92 A.P.C. Road, Calcutta-700009, India.\\
\end{center}
\footnotetext{\noindent
Electronic address:\\
\noindent
$^1$ sanyal\_ ak@yahoo.com \\
Present address: Dept. of Physics, Jangipur College, India - 742213.}
\noindent
\abstract{While studying the viscoelastic flow past an infinite plate with suction and constant heat flux between fluid and plate, Raptis and Tziyanidis gave the solution of a pair of equations for velocity and temperature as functions of distance. They then gave some approximate solutions. This letter shows that the approximations are not justified and presents an exact analytical study..}\\
\maketitle
\flushbottom
\section{Introduction:}

Raptis and Tzivanidis \cite{1} extended the work of Pop and Soundalgekar \cite{1a} by considering a two dimensional viscoelastic fluid past an infinite porous plate with a constant heat flux. The steady flow of an elastico-viscous fluid, also called a Walters' liquid B', is governed by the equations of continuity, momentum and energy. These equations were finally reduced to a pair of equations \cite{1}, namely

\be \label{1a} kf''' + f'' + f' = 0,\ee
\be \label{1b} \Theta '' + \sigma \Theta' = -\sigma E f'^2 - k\sigma E f' f'',\ee
with boundary conditions

\be \label{2} \begin{split} & f = 0,\hspace{0.9 in}\Theta = -1 \hspace{.5 in}\mathrm{as} ~~ \eta = 0,\\& f\rightarrow 1,\hspace{0.87 in}\Theta\rightarrow 0 \hspace{.58 in}\mathrm{as} ~~\eta \rightarrow \infty, \end{split}\ee
upon introduction of the following non-dimensional parameters,
\be \label{3} \begin{split} &  \eta = {y v_0\over \nu}; \hspace{1.67 in} u = U_{\infty} f(\eta)\\&
\sigma = {\rho_\nu c_p\over \lambda}; \hspace{1.6 in} k = {k_0 v_0^2\over \rho \nu^2}\\&
\Theta = {(T - T_{\infty})\lambda v_0 \over q\nu}; \hspace{1.1 in} E = {u_0 U_{\infty}^2\lambda\over q \nu c_p}.\end{split}\ee
In the above, $q$ appearing in the non-dimensional parameters $\Theta$ and $E$ is the heat flux per unit area and all other physical variables have their usual meaning and may be found in \cite{1a}. Frater \cite{2} noted that as $k \rightarrow 0$, $f$ should tend to a Newtonian value, which requires that for small $k$ one can give $f$ approximately by

\be \label{4}  f = 1 - \exp{[-(1 + k)\eta]}.  \ee
Putting equation \eqref{4} in equation \eqref{1b}, Raptis and Tzivanidis obtained an expression for $\Theta$ that satisfied the boundary condition \eqref{2}. Then the authors used these expressions to make explicit numerical calculations with $k = 0.1$ and $k = 0.3$.\\

However in this letter we shall show that equation \eqref{4} leads to different expressions for $f$, depending on whether $k$ is greater than, equal to or less than ${1\over 4}$. Thus, Frater's \cite{2} approximation breaks down completely for $k > {1\over 4}$, which includes the case of $k = 0.3$ studied by Raptis and Tziyanidis \cite{1}. For this reason, we have made an analytical study keeping $k$ completely arbitrary.

\section{Solutions:}

Let $f' = g$, so equation \eqref{1a} becomes

\be  kg'' + g' + g = 0.  \ee
Using the $D$ operator this becomes

\be \label{5}(k D^2 + D + l) g = 0,\ee
So,

\be D = {- 1 \pm (1 - 4k)^{1\over 2}\over 2k}.\ee
Let $\alpha_l$ and $\alpha_2$ be the two roots, so that

\be\label{6}\begin{split}& \alpha_1 + \alpha_2 = -{1\over k},\\&
\alpha_1\alpha_2 = {1\over k}.\end{split} \ee

\subsection{The case with $k < {1\over 4}$:}

The solution of equation \eqref{5} is

\be g = f' = \alpha_1 A e^{\alpha_1 \eta} + \alpha_2 B e^{\alpha_2 \eta},\ee
which on integration gives

\be \label{7} f = A e^{\alpha_1 \eta} + B e^{\alpha_2 \eta} + l + c,\ee
where, we have split the constant of integration into a pair, $l$ and $c$ for convenience, the reason will be clear shortly. Now integrating constants equation \eqref{1a} we get

\be \label{8} k f'' + f' + f = l.\ee
Putting equation \eqref{8} in equation \eqref{1b}

\be \Theta'' + \sigma\Theta' = \sigma E f'(f-l).\ee
On integration one obtains,

\be \Theta' + \sigma\Theta = {\sigma E \over 2}(f - l)^2
 + D',\ee
where $D'$ is the constant of integration. Substituting the value of $f$ from equation \eqref{7}, multiplying both sides by $e^{\sigma n}$ and then integrating again we obtain,

\be \int \big(\Theta e^{\sigma\eta}\big)'d\eta = {\sigma E\over 2}\int e^{\sigma\eta}\big(A e^{\alpha_1 \eta} + Be^{\alpha_2 \eta} +c\big)^2 d\eta + \int D' e^{\sigma\eta} d\eta.\ee
Evaluating the integral we get

\be \begin{split}\label{9} &\Theta = {\sigma E\over 2}\Bigg({A^2\over 2\alpha_1 + \sigma}e^{2\alpha_1 \eta} + {B^2\over 2\alpha_2 + \sigma}e^{2\alpha_2 \eta} + {c^2\over \sigma} + {2A B \exp{[(\alpha_1 + \alpha_2)\eta]}\over \alpha_1 + \alpha_2+ \sigma} + {2Bc\over \alpha_2 +\sigma}e^{\alpha_2\eta}\\&
\hspace{2.6 in}+ {2Ac\over \alpha_1 +\sigma}e^{\alpha_1\eta}\Bigg) + m e^{-\sigma\eta} + {D'\over \sigma}, \end{split} \ee
where m is the constant of integration. Now substituting the values of $f,~ f',~f''$ from equation \eqref{7} in equation \eqref{8}, one obtains

\be A\big(k\alpha_1^2 + \alpha_1 + 1\big)e^{\alpha_1\eta} + B\big(k\alpha_2^2 + \alpha_2 + 1\big)e^{\alpha_2\eta} + c = 0,\ee
which implies $c = 0$. Again applying boundary condition \eqref{2} in equation \eqref{7} we get

\be A + B + l + c = 0; \hspace{0.5 in} l + c = 1,\ee
but since $c = 0$, so finally we obtain

\be\label{10} c = 0;\hspace{0.5 in} l = 1;\hspace{0.5 in} A + B + 1= 0.\ee
Now differentiating equation \eqref{9} and applying the first boundary condition of \eqref{2} and equations \eqref{10} we get

\be\label{11}\begin{split}&
{\sigma E\over 2}\left({2\alpha_1 A^2\over 2\alpha_1 + \sigma} + {2\alpha_2\over 2\alpha_2+\sigma}(A+1)^2 - {2(\alpha_1+\alpha_2)\over\alpha_1+\alpha_2+\sigma}A(A+1)\right) -m\sigma =-1,\\&
m = {1\over \sigma} + A^2 E \left({\alpha_1\over 2\alpha_1+\sigma}+{\alpha_2\over 2\alpha_2+\sigma}-{\alpha_1+\alpha_2\over\alpha_1+\alpha_2+\sigma} \right) + AE\left({2\alpha_2\over 2\alpha_2+\sigma}- {\alpha_1+\alpha_2\over\alpha_1+\alpha_2+\sigma}\right) + {E\alpha_2\over 2\alpha_2 + \sigma},\end{split}\ee
while the application of the second boundary condition of expressions \eqref{2} in equation \eqref{9} gives $D' = 0$. Now applying equation \eqref{10} in equations \eqref{7} and \eqref{9}

\be\label{12}\begin{split}&
f = 1 + A\big(e^{\alpha_1\eta}-e^{\alpha_2\eta} \big) -e^{\alpha_2\eta}\\&
\Theta = {\sigma E\over 2}\left({A^2\over 2\alpha_1 + \sigma}e^{2\alpha_1\eta} + {(A+1)^2\over 2\alpha_2+\sigma}e^{2\alpha_2\eta} - {2A(A+1)\over\alpha_1+\alpha_2+\sigma}\exp{[(\alpha_1+\alpha_2)\eta]}\right) + me^{-\sigma\eta},\end{split}\ee
where $m$ and $A$ are connected by equation \eqref{11}.

\subsection{The case with $k > {1\over 4}$:}

The same procedure as in section (2.1) is applied here, except that $\alpha_1$ and $\alpha_2$ are now both complex; $A$ and $B$ are also complex but $g$ is real. Let $B = A^*$, equation \eqref{10} then gives

\be c = 0, \hspace{0.5 in}l = 1,\hspace{0.5 in} A+ A^* = -1.\ee
Now considering

\be A = a + i b,\hspace{0.5 in}A^* = a - i b,\ee
where both $a$ and $b$ are real, we obtain

\be a = -{1\over 2}, \hspace{0.5 in} A= -{1\over 2} + i b, \hspace{0.5 in}A^* = -{1\over 2} - i b,\ee
and so

\be f = 1 - \Big({1\over 2} - i b\Big)\big(e^{\alpha_1 \eta}- e^{\alpha_2 \eta}\big) - e^{\alpha_2 \eta},\ee
and

\be\Theta = {\sigma E\over 2}\left[{\big(i b - {1\over 2}\big)^2\over 2\alpha_1 +\sigma}e^{2\alpha_1\eta} + {\big(i b + {1\over 2}\big)^2\over 2\alpha_2 +\sigma}e^{2\alpha_2\eta} - {(2ib -1)(2ib+1)\exp{[(\alpha_1 + \alpha_2)\eta]}\over 2(\alpha_1+\alpha_2 + \sigma)} \right] + me^{-\sigma\eta},\ee
where $m$ and $b$ and hence $A$ are connected by equation \eqref{11}. Now, since $f$ and $\Theta$ are both real they can be written in the following form:

\be\label{13} \begin{split} & f = 1 - e^{-{\eta\over 2k}}\left[\cos{\left({(4k -1)^{1\over 2}\over 2k}\eta\right)} - 2b\sin{\left({(4k -1)^{1\over 2}\over 2k}\eta\right)} \right],\\&
\Theta = \Big({\sigma E\over 2}\Big) e^{-{\eta\over k}}\Bigg[\left\{\Big({1\over 4} - b^2\Big)\Big(\sigma-{1\over k}\Big) - {b(4k - 1)^{1\over 2}\over k}\right\}\cos{\left({(4k-1)^{1\over 2}\over k}\eta\right)}\\&
\hspace{0.8 in}+
\left\{\Big({1\over 4} - b^2\Big){(4k - 1)^{1\over 2}\over k}-b\Big(\sigma-{1\over k}\Big)\right\}\sin{\left({(4k-1)^{1\over 2}\over k}\eta\right)}\\&
\hspace{1.1 in}\times\left\{\Big(\sigma -{1\over k}\Big)^2 + {4k-1\over k^2}\right\}^{-1} + {4b^2 + 1\over 2\left(\sigma - {1\over k}\right)} + m e^{-\sigma\eta}\Bigg]\end{split}\ee
where $m$ and $b$ and hence $A$ are connected by equation \eqref{11}.

\subsection{The case with $k = {1\over 4}$:}

Here,
\be g = f' = (A' + B' \eta)e^{-2\eta}.\ee
Integrating, we get

\be \label{14} f = - \left({B'\over 2} + A' + B'\eta\right){e^{-2\eta}\over 2}+ D + l',\ee
where $D$ and $l’$ are constants of integration. Substituting equations \eqref{8} and \eqref{14} in equation \eqref{1b}, multiplying both sides by  $e^{\sigma\eta}$ and integrating we get

\be \big(\Theta e^{\sigma\eta}\big)' = {\sigma E\over 2} e^{\sigma \eta}\left[D - \left(A' + B'\eta+{B'\over 2}\right){e^{-2\eta}\over 2}\right]^2 + D'' e^{\sigma \eta},\ee
where $D''$ is a constant of integration. This on integration yields

\be \label{15}\begin{split} & \Theta = {\sigma E\over 2}\Bigg\{{D^2\over \sigma} - {(2A'+B')D\over 2(\sigma -2)}e^{-2\eta} + {(2A' + B')^2\over 16(\sigma - 4)}e^{-4\eta} - {B'D\eta \over \sigma -2}e^{-2\eta} + {B'D\over (\sigma-2)^2} e^{-2\eta}\\&
+ {(2A'+B')\over 4(\sigma-4)}B'\eta e^{-4\eta} -{(2A'+B')B'\over 4(\sigma-4)^2}e^{4\eta} + {B'^2\eta^2\over 4(\sigma -4)}e^{-4\eta}
-{B'^2\eta\over 2(\sigma-4)^2}e^{-4\eta} + {B'^2\over 2(\sigma-4)^3}e^{-4\eta}\Bigg\}\\& + ne^{\sigma\eta} + {D''\over \sigma}.\end{split}\ee
Now, substituting the values of $f$, $f’$, $f''$ from equation \eqref{14} in equation \eqref{8}, we get:

\be \left[2kB' -\left({B'\over 2} + A' + B'\eta\right)\left(2k - {1\over 2}\right) - {B'\over 2}\right]e^{-2\eta} + D = 0,\ee
which implies $D = 0$. So equation \eqref{15} becomes

\be \label{16}\begin{split} & \Theta = {\sigma E\over 2}\Bigg[{(2A' + B')^2\over 16(\sigma - 4)} + {(2A' + B')B'\over 4(\sigma - 4)}\eta - {(2A' + B')B'\over 4(\sigma - 4)^2} + {B'^2\eta^2\over 4(\sigma - 4)}  - {B'^2\eta\over 2(\sigma - 4)^2} + {B'^2\over 2(\sigma - 4)^3}\Bigg]e^{-4\eta}\\&\hspace{2.4 in} + n e^{-\sigma\eta} + {D'\over \sigma}. \end{split}\ee
Now applying the boundary conditions \eqref{2} in equation \eqref{14} gives:

\be -{1\over 2}\left(A' + {B'\over 2}\right) + l' = 0,\hspace{0.4 in} \because ~ D = 0, \ee
or,

\be  2A' + B = 4l'; \hspace{0.4 in} l' = 1.\ee
So, finally we get

\be \label{17} A' + B' = 4;\hspace{0.5 in} l' = 1.\ee
Now differentiating equation \eqref{16} and applying the first boundary condition of \eqref{2} gives

\be{\sigma E\over 2}\left[{(2A' + B')B'\over 4(\sigma - 4)} + { B'^2\over 2(\sigma - 4)^2}\right] - 2\sigma E\left[{(2A' + B')^2\over 16(\sigma - 4)} - {B'^2\over 2(\sigma - 4)^3}  - {(2A' + B')B'\over 4(\sigma - 4)^2}\right]- n\sigma + 1 = 0.\ee
Also applying equation \eqref{17}, we obtain,

\be \label{18} n = {E\over 2}\left[{B'\over \sigma -4} - {B'^2\over 2(\sigma -4)^2}\right] - 2E \left[{1\over \sigma -4} - {B'\over (\sigma -4)^2} + {B'^2\over 2(\sigma -4)^3}\right] + {1\over \sigma}.\ee
Applying the second boundary condition  \eqref{2} in equation \eqref{16} we get $D” = 0$. Hence using equations \eqref{17} and \eqref{18}, equations \eqref{14} and \eqref{16} become

\be \label{19}\begin{split}&
f = 1 - {1\over 2}(B'\eta + 2) e^{-2\eta},\\&
\Theta = {\sigma E\over 2(\sigma -4)}\left[{B'^2\over 2(\sigma - 4)^2} - {B'^2\eta \over 2(\sigma - 4)} - {B'^2\eta^2\over 4}- {B'\over \sigma - 4} + B'\eta +1 \right]e^{-4\eta} + n e^{-\sigma \eta}, \end{split}\ee
where $n$ and $B$ are related by equation \eqref{18}.

\section{Discussion:}

In summary, solutions of equations \eqref{1a} and \eqref{1b} subject to boundary conditions \eqref{2} are given by equation \eqref{12} for $k < {1\over 4}$, equation \eqref{13} for $k > {1\over 4}$ and by equation \eqref{19} for $k = {1\over 4}$. Note that in all the three cases, there is one unspecified constant. However, equations \eqref{12} and \eqref{13} are such that

\be \lim_{k\rightarrow {1\over 4} - 0}{f} = \lim_{k\rightarrow {1\over 4} + 0} {f} = \lim_{k\rightarrow {1\over 4}} {f}.\ee
If we make the physical assumption that

\be \lim_{k\rightarrow {1\over 4}} {f} = f, \hspace{0.2 in} \mathrm{at} \hspace{0.2 in} k = {1\over 4},\ee
it is easy to see that one must have $B’ = 0$ and hence $A ’ = 2$ in equation \eqref{19} for $k = {1\over 4}$. For $k < {1\over 4}$, if following Frater \cite{2}, we assume that as $k \rightarrow 0$, $f$ tends to Newtonian values, then comparing equations \eqref{4} and \eqref{12} we get $A = - 1$ in equation \eqref{12} for $k < {1\over 4}$. For $k > {1\over 4}$ however, the present authors do not have any idea as to how the unspecified constant $b$ in equation \eqref{13} can be specified. We further note that $f$ takes completely different expressions for $k < {1\over 4}$ and $k > {1\over 4}$. Hence Frater’s approximation \cite{2} completely breaks down for $k > {1\over 4}$. Therefore  use of Frater’s approximation \cite{2} by Raptis and Tzivanidis’ \cite{1} for $k = 0.3$ is unjustified.

\end{document}